# Observation of magnon-mediated current drag in Pt/yttrium iron garnet/Pt(Ta) trilayers

Junxue Li[1*], Yadong Xu[1*], Mohammed Aldosary[1], Chi Tang[1], Zhisheng Lin[1], Shufeng Zhang[2], Roger Lake[3], and Jing Shi[1]

1. Department of Physics and Astronomy, University of California, Riverside, CA 92521
2. Department of Physics, University of Arizona, Tucson AZ 85721
3. Department of Electrical and Computer Engineering, University of California, Riverside CA 92521

Pure spin current, a flow of spin angular momentum without flow of any companying net charge, is generated in two common ways. One makes use of the spin Hall effect in normal metals (NM) with strong spin-orbit coupling, such as Pt or Ta. The other utilizes the collective motion of magnetic moments or spin waves with the quasi-particle excitations called magnons. A popular material for the latter is yttrium iron garnet, a magnetic insulator (MI). Here we demonstrate in NM/MI/NM trilayers that these two types of spin currents are interconvertible across the interfaces, predicated as the magnon-mediated current drag phenomenon. The transmitted signal scales linearly with the driving current without a threshold and follows the power-law $T^n$ with *n* ranging from 1.5 to 2.5. Our results indicate that the NM/MI/NM trilayer structure can serve as a scalable pure spin current valve device which is an essential ingredient in spintronics.

*These two authors contributed equally.

Corresponding author email: jing.shi@ucr.edu

# Introduction

There has been intense research interest in pure spin current transport in both conducting and insulating materials. Whether by spin pumping[1-4], spin Seebeck effect (SSE)[5-8], or spin Hall and inverse spin Hall effects (SHE/ISHE)[9-15], pure spin current generation and detection are typically accomplished in bilayers consisting of a magnetic and a non-magnetic layer[2-4]. Either generation or detection, rarely both, is done by electrical means. Kajiwara et al[16]. first demonstrated a lateral all-electrical device in which an electrical signal can transmit through yttrium iron garnet (YIG) over a macroscopic distance (~1 mm). The response (~ 1 nV) is highly nonlinear with a threshold in driving current, which was interpreted as the critical value for the spin transfer torque[17,18] induced coherent precession of YIG magnetization. The effect also suggests that spin currents of two different types, i.e. electronic[9-15] and magnonic[19-22], are inter-convertible at the interfaces. However, the phenomenon has not been reproduced in a similar planar geometry.

Zhang et al.[23,24] predicted a phenomenon with a different origin in sandwich structures, analogous to the conventional spin-valve device for spin-polarized charge currents[25,26]. Rather than exciting a coherent precession of magnetization, at one interface, electrons in the NM create/annihilate magnons in the MI. The individual magnon creation/annihilation does not need to overcome any threshold; therefore, the interconversion takes place at any current. Due to the long magnon decay length in MI[27], this effect couples two remote electrical currents, which is called the magnon mediated current drag. Moreover, the on and off states are controlled by the relative orientation between the polarization of the spin current and MI magnetization. In small lateral devices similar to that of Kajiwara's, Cornelissen et al.[28] reported both $\omega$- and $2\omega$ nonlocal current responses in the ac measurements ($\omega$ is the frequency of ac current). The $\omega$-response indicates a linear nonlocal current relation which apparently does not share the same origin as that of the effect reported by Kajiwara. Nonlocal magnetoresistance was also reported in similar device geometry[29]. However, it is unclear whether the reported effect indeed arises from the microscopic mechanism predicted by Zhang et al. An objective of this work is to experimentally establish the current drag effect in the sandwich geometry proposed by Zhang et al.

In this work, we first successfully fabricate Pt/YIG/Pt(Ta) trilayer structures using pulsed laser deposition in combination with sputtering and post-growth annealing. Below 220 K, the trilayers are sufficiently insulating and we observe nonlocal responses when an in-plane magnetic field is swept or rotated. We find that the polarity of the nonlocal signal is opposite to each other between Pt/YIG/Pt and Pt/YIG/Ta, indicating the spin current origin of the effect. By investigating the effect at low temperatures, we show that the power-law dependence of the nonlocal signal is consistent with the prediction for the magnon-mediated current drag effect.

**Results**

**Nonlocal device structure**. Our NM/MI/NM trilayer device structure is schematically shown in figures 1a and 1b. The MI is a thin YIG film sandwiched by either two identical NM films or dissimilar NM films. Via the SHE, a charge current ($J_{injected}$) generates a pure spin current flowing in the z-direction with the spin polarization ($\sigma$) parallel to the y-direction. The conduction electrons in the bottom NM interact with the localized moments of the MI via the s-d exchange interaction at the interface, resulting in the creation or annihilation of magnons in the MI accompanied by spin-flips of conduction electrons in the bottom NM layer. Due to the nature of the s-d exchange interaction[24,30], i.e. $H_{sd} = -J_{sd} \sum \sigma \cdot M$, where $J_{sd}$ is exchange coupling strength, when the magnetization of the MI (**M**) is collinear with **σ**, magnons are created or annihilated depending on whether **M** is parallel or anti-parallel to **σ**. As such, the interaction creates a non-equilibrium magnon population and spin accumulation in the MI which drives magnon diffusion. The excess/deficient magnons are then converted to a spin current in the top NM layer by the reverse process, which is converted to a charge current ($J_{induced}$) again in the top NM layer via the ISHE. When **σ** ⊥ **M**, there is no non-equilibrium magnon population or spin accumulation and the spin current is absorbed by the MI. Consequently, there is no induced spin or charge current in the top NM layer (figure 1b). Remarkably, one can switch on and off the magnon creation/annihilation process by controlling the relative orientation between **M** and **σ**. Conceptually, this structure functions as a valve for pure spin current.

**Field-dependent nonlocal response in trilayer devices.** Since both Pt and Ta have strong spin-orbit coupling with opposite signs in their spin Hall angle[31-33], we have fabricated three Pt(5 nm)/YIG(80 nm)/Pt(5 nm) devices and two Pt(5 nm)/YIG(80 nm)/Ta(5 nm) reference samples, which were deposited on (110)-oriented $Gd_3Ga_5O_{12}$ (GGG) substrates (see Methods for fabrication details). The inset of figure 2c shows an optic image for a GGG/Pt/YIG/Pt device. As illustrated in figure 2b, the bottom Pt layer is used to inject current $I_b$, while the top layer, either Pt or Ta, functions as a detector to measure the induced current or the nonlocal voltage $V_{nl}$. An in-plane $H$ is either swept in a fixed direction or rotated with a continuously varying angle of $\theta$ measured from the y-direction. We find that the high-quality YIG/Pt interface is essential to the observation of the spin current transmission. As shown in figure 2a, the morphology of a YIG film tracks the atomically flat terraces of the GGG (110) surface[34] in spite of a layer of Pt in between. The excellent interface quality is verified by both the SSE[5] (see Supplementary Note 5 and Supplementary Fig. 6) and the spin Hall magnetoresistance (SMR)[35,36] (see Supplementary Note 3 and Supplementary Fig. 3) in the same devices. The 80 nm thick YIG films are nearly insulating but have small leakage at high temperatures. However, the resistance between the top and bottom NM layers increases exponentially as the temperature ($T$) decreases (see Supplementary Note 2 and Supplementary Fig. 2), and exceeds $20\,G\Omega$ at and below 220 K. Therefore, all nonlocal measurements were performed below 220 K to avoid any parasitic signal from the small leakage current. In $V_{nl}$, we remove a non-zero background signal that exists even at $I_b$=0.

Figures 2c and 2d plot the field dependence of $V_{nl}$ at 220 K. When H is swept along $I_b$, i.e. $\theta = 90^o$ (figure 2c), $V_{nl}$ is a constant at $I_b$= 0 (red). However, at $I_b$= +1.5 mA, $V_{nl}$ shows a clear hysteresis with two positive peaks tracking the coercive fields of the YIG film, indicating that $V_{nl}$ is closely related to the magnetization state of YIG. As the $I_b$ is reversed, $V_{nl}$ also reverses the sign. In principle, a sign reversal can occur if there is a finite leakage current flowing in the top layer. Through the magnetoresistance, this current can produce a hysteretic voltage signal. Estimating from the leakage current, we find that the relative change in $V_{nl}$ due to this effect is at least three orders of magnitude smaller than the observed nonlocal voltage signal (see Supplementary Note 4 and Supplementary Figs. 4 and 5). Therefore we exclude the leakage current as the source of the nonlocal signal. Note that $V_{nl}(\pm 1.5$ mA) is the same as $V_{nl}(0$ mA) at the saturation state (H >200 Oe) when $\sigma \perp M$, suggesting magnon creation/annihilation is totally suppressed. For the field sweeps with $\theta = 0^o$ (figure 2d), $\sigma$ is collinear with $M$ at high fields, interface magnon creation/annihilation results in a full current drag signal. Clearly, $V_{nl}(+1.5$ mA) is different from $V_{nl}(0$ mA) at the saturation fields and reverses the sign when $I_b$ reverses. It is interesting to note that $V_{nl}(\pm 1.5$ mA) differ from $V_{nl}(0$ mA) at the coercive fields. One would expect them to be the same since the average magnetization should point to the x-direction at the coercive fields, which would correspond to the saturation states for $\theta = 90^o$ in figure 2c. This discrepancy can be explained by the multi-domain state of YIG in which the actual $M$ is distributed over a range of angles around $\theta = 90^o$, and the collinear component of $M$ turns on the magnon channel and yields a nonzero $V_{nl}$. In order to investigate the phenomenon in the single-domain state, we perform the following experiments.

**Angle-dependent nonlocal response of single-domain YIG.** Figure 3a presents $V_{nl}$ in GGG/Pt/YIG/Pt as a function of $\theta$ between $M$ and $\sigma$ at 220 K, as illustrated in figure 2b. The 80 nm (110)-oriented YIG grown on Pt has a well-defined uniaxial anisotropy with an anisotropy field < 200 Oe. The applied magnetic field (1000 Oe) is sufficiently strong not only to set YIG into a single-domain state, but also to rotate $M$ with it. For all positive $I_b$ (solid symbols), $V_{nl}$ exhibits maxima at $\theta = 0°$ and 180° ($M$ collinear with $\sigma$), but minima at $\theta = 90^o$ and $270^0$ ($M \perp \sigma$). $V_{nl}$ changes the sign as $I_b$ is reversed (empty symbols). At $\theta = 90^o$ and $270^0$, the nonlocal signal for $\pm I_b$ coincides with $V_{nl}$(0 mA), further validating that the spin current is in the off state when $\mathbf{M} \perp \mathbf{\sigma}$. Similar angular dependent measurements are also taken on a GGG/Pt/YIG/Ta device and the results are depicted in figure 3d. For the same measurement geometry and the same polarity of $I_b$, we find that $V_{nl}$ of GGG/Pt/YIG/Ta has the opposite sign to that of GGG/Pt/YIG/Pt, which is just expected from the opposite signs in their spin Hall angle. The $V_{nl}$ sign difference here is another piece of critical evidence for the magnon mediated mechanism, as opposed to other extrinsic ones such as leakage.

An interesting feature to note here is that $V_{nl}$ at $\theta = 0^o$ and $180^o$ shows a slight but reproducible difference which is independent of the current polarity but increases with the increasing magnitude of $I_b$. We attribute this phenomenon to the SSE contribution since the joule heating in the bottom Pt layer unavoidably generates a small vertical temperature gradient, which in turn launches an upward spin current in YIG entering the top Pt (or Ta) layer. As $\mathbf{M}$ reverses, so does the spin polarization, which consequently produces two different SSE signal levels between $\theta = 0^o$ and $180^o$. Combining these two effects, we can fit the angular dependence data using:

$$V_{nl} = V_0 + V_{SSE} \cos\theta + V_{Drag} \cos^2\theta \qquad (1)$$

where $V_0$ is an offset voltage insensitive to the magnetization orientation, $V_{SSE}$ is the SSE voltage amplitude, and $V_{Drag}$ represents the amplitude of the current drag signal. The solid red curves in figures 3a and 3d fit the experimental data remarkably well, and the extracted fitting results are plotted in figures 3b and 3e for GGG/Pt/YIG/Pt and GGG/Pt/YIG/Ta devices, respectively. Two conclusions can be evidently drawn from these results. First, the magnitude of the current drag signal (red cycles) scales linearly with the driving current, i.e. $V_{Drag} \propto I_b$. This is in stark contrast with the highly nonlinear behavior[16]. Second, the weak current dependence of the SSE contribution follows $V_{SSE} \propto I_b^2$ (as shown in figures 3c and 3f), which is characteristic of thermoelectric effects. Compared with usual bilayers, trilayer structures may have an enhanced SSE contribution due to the presence of the second heavy metal layer that draws an extra heat-driven spin current. Carefully designed experiments are needed to separate this effect.

**Temperature dependence of nonlocal responses.** According to Zhang, et al.[24], the temperature dependence of the injection interface spin convertance $G_{em}$ is $\left(\frac{T}{T_c}\right)^{3/2}$, where $T_c$ is the Curie temperature of the MI; for the detection interface, the spin convertance $G_{me}$ is proportional to $\frac{T}{T_F}$, where $T_F$ is the Fermi temperature of the NM layer. In the most simplified picture which is strictly applicable only for very thick films, the current drag signal should be proportional to the product of the two spin current convertances, i.e. $V_{Drag} \propto G_{em} \cdot G_{me} \propto T^{5/2}$. The representative angular dependence measurements below 220 K are shown in figures 4a and 4c for GGG/Pt/YIG/Pt and GGG/Pt/YIG/Ta devices, respectively. For both samples, $I_b$ is set at +2 mA and $H$ is held at 1000 Oe. The magnitude of the current drag signal decreases progressively with decreasing temperature for both devices. By fitting $V_{nl}$ using Equation (1), we extract the magnitude of $V_{Drag}$ and $V_{SSE}$ shown in figures 4b and 4d. Apart from the expected sign difference, the magnitude of $V_{Drag}$ in both devices monotonically decreases with decreasing temperature. In fact, both data sets can be well fitted by a power-law $V_{Drag} = V_{Drag}^0 T^n$ (red solid curves in figures 4b and 4d), where $V_{Drag}^0$ is a pre-factor. The extracted exponent $n$ is 2.21 for GGG/Pt/YIG/Pt and 1.88 for GGG/Pt/YIG/Ta, falling in the range between 1.5 and 2.5. It should be pointed out that the full picture described in Ref. 24 actually contains other quantities that have weak temperature dependence. The deviation of the exponent from 2.5 is fully expected if these factors are considered. On the other hand, the $V_{SSE}$ is found to be relatively insensitive to temperature, suggesting a completely different mechanism.

In conclusion, we experimentally establish the magnon mediated current drag effect in NM/MI/NM trilayer structures by investigating the field-, angle-, current- and temperature-dependences of the nonlocal signal. The spin information carried by conduction electrons and magnons in different materials can be interconverted at the interfaces. Such structures can serve as pure spin current valve devices since rotating the in-plane magnetization of the MI by 90$^o$ provides a digital on-off switch of the spin current. Furthermore, such structures also provide analog functionality since rotating the in-plane magnetization of the MI provides analog sinusoidal modulation of the spin current. Due to the extremely low damping in the MI, transmission of the pure spin currents can occur over relatively long distances providing the functionality of a pure spin interconnect.

# Methods

**Fabrication** A Hall bar was first defined in the photo-resist layer on a (110)-oriented single crystalline GGG substrate using photolithography with the channel width of 20 μm and the length of 300 μm between two voltage electrodes. Then the bottom Pt layer was deposited on the open Hall bar area by dc magnetron sputtering. During sputtering, argon pressure was 5 mTorr, substrate temperature was kept at 300 K, and the dc sputtering power was 37.5 W. The deposition rate of Pt was 0.77 Å/s and the Pt layer thickness was 5 nm. After liftoff, an 80 nm thick YIG film was deposited at 450 $^o$C with $O_2$ pressure of 1.5 mTorr by pulsed laser deposition (PLD) to cover the surface of the entire sample. The as-grown YIG film became crystalized and magnetized after rapid thermal annealing (RTA) between 800 to 850 $^o$C for 200 s. We had explored a range of growth temperatures, different annealing conditions, different PLD rates, and YIG film thicknesses and had experienced many difficulties such as YIG film cracking, peeling off for thicker YIG films, conducting, non-magnetic, etc. before we identified the working window. The magnetic properties of YIG were investigated by vibrating sample magnetometer (VSM) (see Supplementary Note 1 and Supplementary Fig. 1). The surface morphology of YIG was monitored by atomic force microscopy (AFM). Clear atomic terraces were observed and the root-mean-square (RMS) roughness on terraces was ~0.14 nm, indicating a very flat YIG surface. The top Pt and Ta patterns were defined using standard e-beam lithography, followed by magnetron sputtering deposition and lift-off procedures. Prior to Pt (or Ta) deposition, 60 s argon ion milling was used to remove any polymer residues from the YIG surface. The deposition conditions for top Pt and Ta were the same as those for bottom Pt. The top Pt and Ta strips are 2 μm in width, and 70 μm and 60 μm in length, respectively. In order to generate a vertical temperature gradient in separate longitudinal Spin Seebeck effect measurements which we conducted after the nonlocal measurements were finished, we deposited 300 nm thick $Al_2O_3$ on top and then Cr(5 nm)/Au(50 nm), which serves as a heater for SSE measurements.

**Transport measurement** For all transport measurements, the current was fed to the devices using a Keithley 2400 dc current source, and the voltage was measured by a Keithley 2182A nano-voltmeter. The field dependence measurements were carried out using a closed cycle system, while the angular dependent measurements were performed by a physical property measurement system (PPMS) equipped with a rotatory sample holder. For the non-local measurements, the excitation current in the bottom Pt is usually no more than 2 mA; for the local magnetoresistance measurements, the current applied in top Pt and Ta was 1 µA, while the current used in bottom Pt is 10 µA. For the SSE measurements, the heating current applied in the top Au layer is 30 mA. In all measurements, extra precaution was taken to ensure the correct polarity of both current and voltage.

**Acknowledgements** We would like to thank Y. Barlas, N. Amos, G. Yin, and S.S. Su for the technical assistance and fruitful discussions. Sandwich structure growth, device fabrication and characterization, electrical transport and thermoelectric measurements, and data analyses at UCR were supported as part of the SHINES, an Energy Frontier Research Center funded by the U.S. Department of Energy, Office of Science, Basic Energy Sciences under Award # SC0012670. S.F. Zhang was supported by NSF-ECCS-1404542.

**Author Contributions** J.S. designed and supervised the project. J.X.L. and Y.D.X. performed device fabrication and transport measurements with the help of Z.S.L.. M.A. performed the pulsed laser deposition of the sandwich structures and characterization with the help of C.T.. S.F.Z. and R.L. contributed to data analyses. J.X.L. and J.S. wrote the manuscript and all the authors contributed to the final version of manuscript.

**Figures**

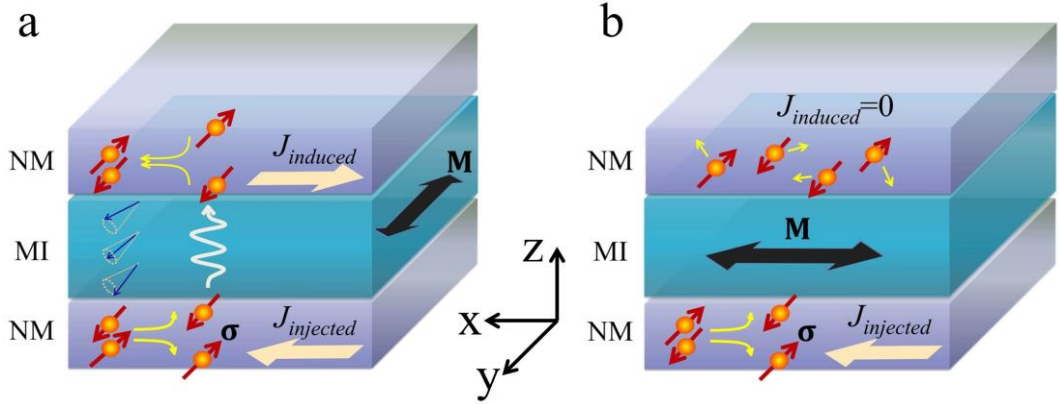

**Figure 1. Schematic illustration of spin current valve. a,** the transmission of spin current is switched on. Magnetization (**M**) of magnetic insulator (MI) oriented collinearly with the spin polarization **σ** (//y) of the pure spin current in the bottom normal metal (NM) layer generated by the spin Hall effect with an electric current $J_{injected}$. The spin-flip scattering of conduction electrons at the bottom NM/MI interface can create (**M** ∥ −**σ**) or annihilate (**M** ∥ **σ**) magnons. A non-equilibrium magnon population extends to the top MI/NM interface, and the spin angular momentum carried by magnons is transferred to conduction electrons in the top NM layer. The pure spin current flowing perpendicular to the NM layer is then converted to a charge current ($J_{induced}$) via the inverse spin Hall effect. **b,** the transmission of spin current is switched off. **M** perpendicular to the spin polarization **σ** of the spin current. In this geometry, the s-d exchange interaction between conduction electrons and local magnetic moments does not excite magnons in the MI. Consequently, there is no spin accumulation at the top MI/NM interface or induced charge current in the top NM layer.

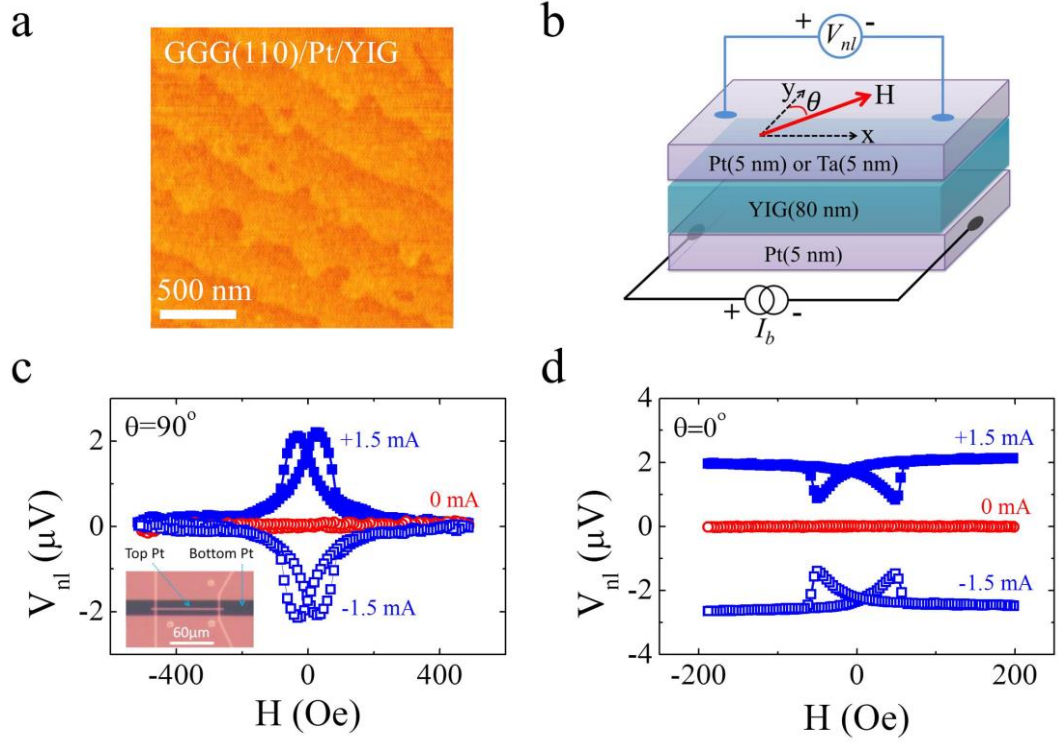

**Figure 2. Mesaurement geometry and field-dependent nonlocal signal. a,** AFM image of 80 nm YIG film grown on 5 nm Pt. **b,** Schematic illustration of the experimental set-up. $I_b$ is the current applied to the bottom Pt layer, and $V_{nl}$ is the nonlocal voltage measured at top layer along the $I_b$ direction. The applied, in-plane magnetic field $H$ makes an angle $\theta$ with the y-axis which is in plane and perpendicular to the current direction (x). **c,** The field dependence of the nonlocal signal for $H$ along $I_b$, i.e., $\theta = 90^o$. The inset shows the optical image of GGG/Pt/YIG/Pt device. **d,** The field dependent nonlocal signal with $H$ perpendicular to $I_b$, i.e., $\theta = 0^o$. In both **c** and **d**, solid (empty) blue squares and empty red cycles represents $V_{nl}$ for +1.5 mA (-1.5 mA) and 0.0 mA bottom current, respectively.

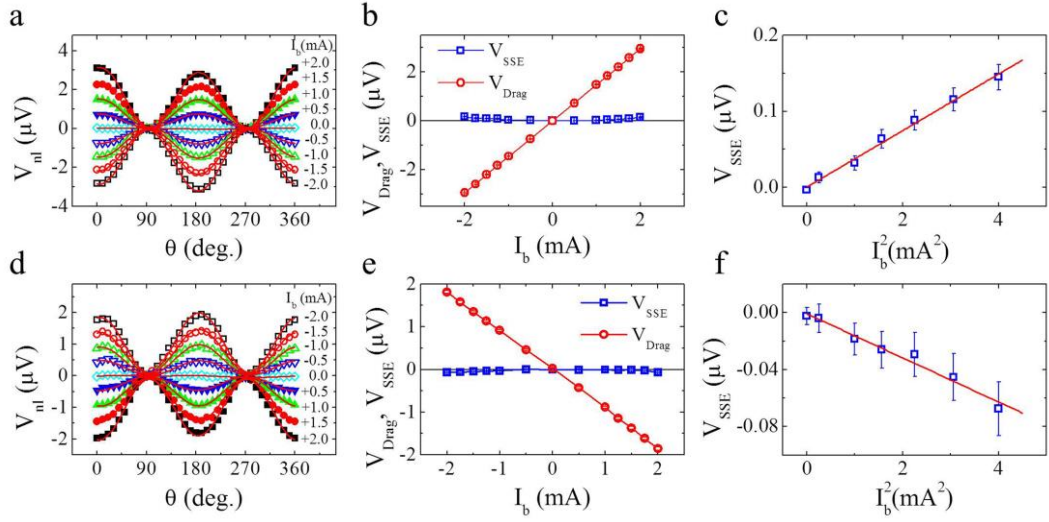

**Figure 3. Angular and current dependence of nonlocal signal**. **a** and **d** are angular dependence of nonlocal signal at different currents in the bottom Pt layer for GGG/Pt/YIG/Pt and GGG/Pt/YIG/Ta, respectively. The magnetic field is fixed at 1000 Oe and rotated in plane. In **a** and **d**, solid symbols indicate positive $I_b$, and empty symbols indicate negative $I_b$. Red curves are the fits using equation (1). **b** and **e** show the $I_b$-dependence of the current drag signal ($V_{Drag}$) and the spin Seebeck signal ($V_{SSE}$) for GGG/Pt/YIG/Pt and GGG/Pt/YIG/Ta, respectively. **c** and **f** are the spin Seebeck signal as a function of $I_b^2$ for GGG/Pt/YIG/Pt and GGG/Pt/YIG/Ta, respectively, the red curves are the linear fits. The error bars are from fitting using equation (1) in **a** and **d**.

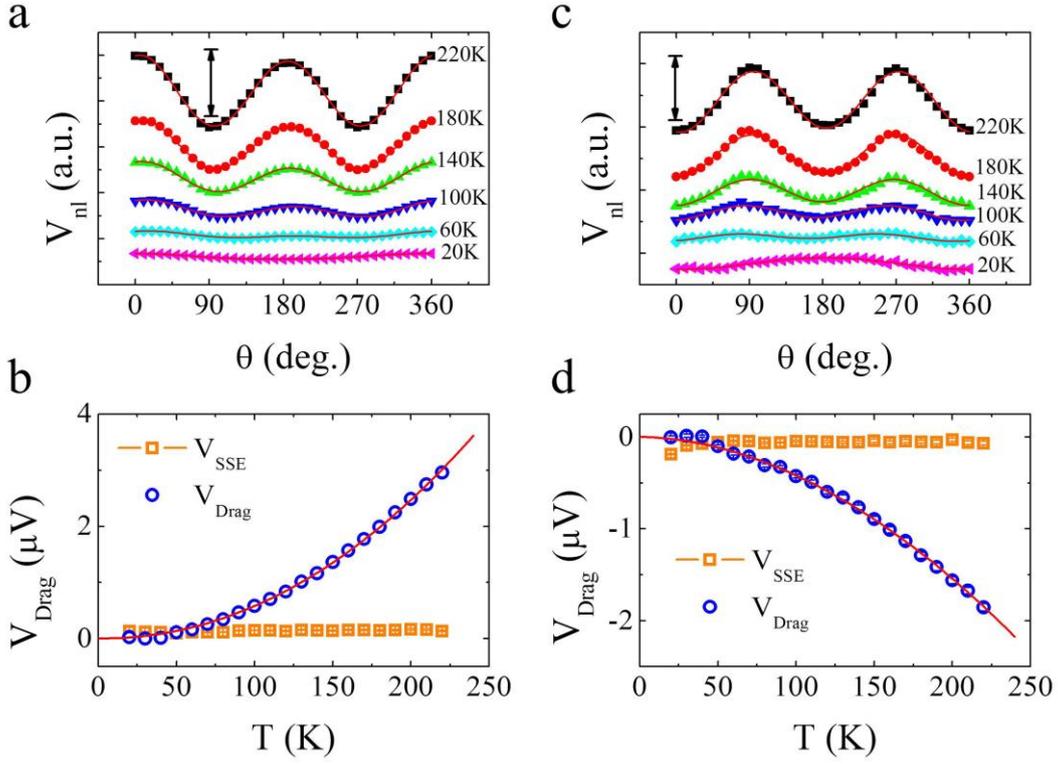

**Figure 4. Temperature dependence of nonlocal signal**. **a** and **c** are angular dependence of the nonlocal signal at different temperatures for GGG/Pt/YIG/Pt and GGG/Pt/YIG/Ta, respectively. During the measurements, $I_b$ was fixed at +2 mA. The curves are vertically shifted for clarity, the black arrows in **a** and **c** represent the magnitude scale of 2.80 μV and 1.98 μV, respectively. Red solid curves in **a** and **c** are the fits using equation (1). **b** and **d** are the temperature dependence of the extracted current drag signal ($V_{Drag}$) and spin Seebeck signal ($V_{SSE}$) for GGG/Pt/YIG/Pt and GGG/Pt/YIG/Ta, respectively. Red solid curves in **b** and **d** are the fits using $V_{Drag} = V_{Drag}^0 T^n$, here, n=2.21 for the GGG/Pt/YIG/Pt device, and n=1.88 for the GGG/Pt/YIG/Ta device.